\documentclass[aps, amssymb, amsmath, superscriptaddress, prl, twocolumn]{revtex4-2}

\usepackage[colorlinks,linkcolor=blue,citecolor=blue,urlcolor=blue]{hyperref}

\usepackage{graphicx}
\usepackage{color}
\usepackage{amsmath}
\usepackage{enumitem}
\usepackage{amssymb}
\usepackage{hyperref}
\usepackage{cancel}
\usepackage{ulem}
\usepackage{multirow}

\usepackage{pifont}
\usepackage{physics}
\usepackage[T1]{fontenc}

\newcommand{\be}{\begin{equation}}
	\newcommand{\ee}{\end{equation}}
\newcommand{\bea}{\begin{eqnarray}}
	\newcommand{\eea}{\end{eqnarray}}

\renewcommand{\Im}{{\rm \, Im\,}}

\renewcommand{\vec}[1]{{\boldsymbol #1}}

\renewcommand\vec[1]{\ensuremath\boldsymbol{#1}} 

\usepackage{amsfonts, relsize, color}
\usepackage{graphicx}
\usepackage{color}
\usepackage{comment}

\usepackage{xcolor}

\usepackage{booktabs} 
\begin{document}
	
	\title{Projected amorphous topological insulators}

\author{Archisman Panigrahi}
\email{archi137@mit.edu}
\affiliation{Department of Physics, Massachusetts Institute of Technology, Cambridge, Massachusetts 02139, USA}

\author{Bitan Roy}
\email{bitan.roy@lehigh.edu}
\affiliation{Department of Physics, Lehigh University, Bethlehem, Pennsylvania 18015, USA}

\begin{abstract}
We introduce projected amorphous topological insulators (PATIs) that despite containing only a fraction ($x$) of otherwise randomly selected {\it disconnected} sites of a parent crystal, feature a quantized global topological invariant ({\it strong} PATIs) when its effective Hamiltonian is constructed by integrating out the residual sites of the original lattice. Within the topological regime of the parent model, additionally, there exists a critical $x$ below which such systems foster a {\it fragile} PATI that only supports quantized local topological marker on a small fraction of sites therein. Both strong and fragile PATIs accommodate in-gap modes near the boundary. The system also hosts a normal insulator in the entire trivial parameter regime of the lattice-based model for any $x$, devoid of gapless boundary modes, for which both global topological invariant and local topological marker vanish. We demonstrate these {\it possibly} generic outcomes by starting with a parent square lattice-based model for time-reversal symmetry breaking insulators. Then the strong-to-fragile PATI quantum phase transition, tunable via $x$, is characterized by the {\it mean} correlation length exponent $\nu \in (1.00,1.46)$.   
\end{abstract}

\date{\today}

\maketitle

{\it Introduction}.~The catalog of insulators can be divided into two broad categories, topological and trivial or normal insulators~\cite{Hasan2010, Qi2011, Chiu2016, Fu2006, Bernevig2006, Roy2009, Ryu2010}. In the former family, electronic Bloch bands invert near high-symmetry point(s) in the Brillouin zone that manifest via a quantized topological invariant, resulting in robust gapless boundary modes; a phenomenon commonly known as the bulk-boundary correspondence. The requisite band inversion can take place at the $\Gamma$ point, residing at the center of the Brillouin zone or at finite momentum point(s) therein, existence of which crucially depends on discrete translational and rotational symmetries of the underlying crystal~\cite{Fu2011, Slager2012, Shiozaki2014, bradlyn2017, Po2017, Fang2019}. Topological insulators from these two families are thus named translationally inert and active, respectively. It is, therefore, of paramount fundamental and practical (such as in low-power logic devices) importance to unveil the possibility of realizing topological insulators in the absence of any crystalline order, such as on amorphous lattices, where the sites are randomly distributed.

Somewhat exclusively and perhaps expectedly, so far, all the studies have found footprints of only translationally inert topological insulators on amorphous lattices~\cite{Agarwala2017, Mitchell2018, Yang2019, Costa2019, Zhou2020, Agarwala2020, Marsal2020, Wang2021, Li2021, Tao2023, Corbae2023, Manna2024}. Here we circumvent this limitation in terms of {\it projected amorphous topological insulators} (PATIs), exemplified in two dimensions, from an otherwise general principle of construction that should be operative in any dimension, and for any symmetry class and parent crystal. 

\begin{figure}[b!]
    \centering
    \includegraphics[width=0.90\linewidth]{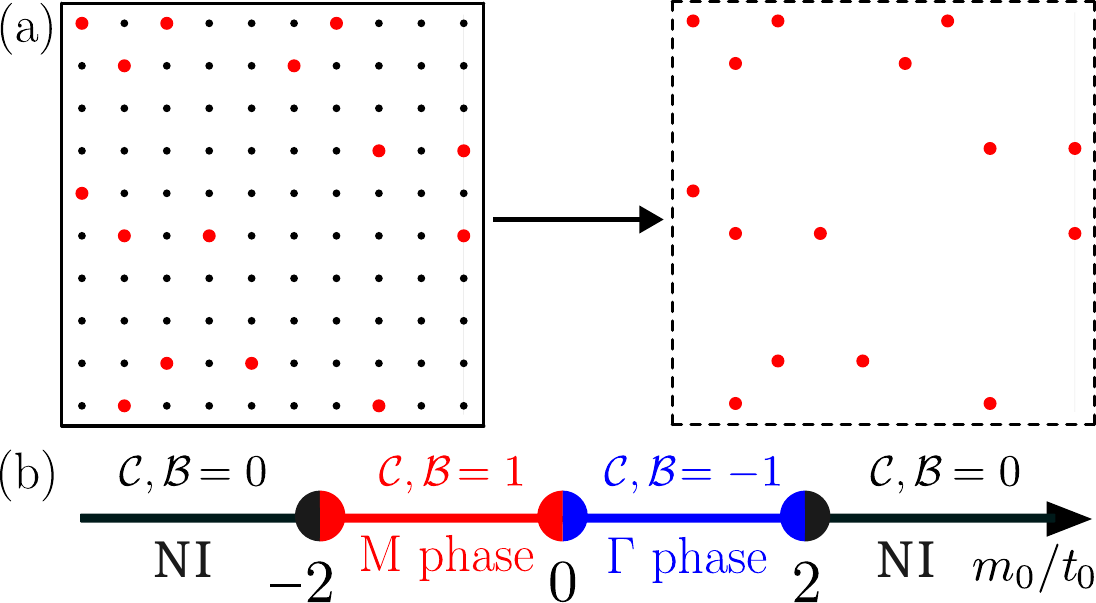}
    \caption{(a) A schematic construction of a two-dimensional amorphous lattice (right) from a randomly selected small fraction ($x$) of sites (red) of the parent square lattice (left). (b) Phase diagram of the Qi-Wu-Zhang model supporting translationally inert $\Gamma$ and active ${\rm M}$ phases (topological), characterized by distinct quantized Chern number (${\mathcal C}$) or Bott index (${\mathcal B}$). For normal insulators (NIs), ${\mathcal C}={\mathcal B}=0$.}
    \label{fig:lattice}
\end{figure}

{\it Key results}.~To construct PATIs, we identify a fraction ($x$) of randomly positioned sites from an underlying crystal, see Fig.~\ref{fig:lattice}(a). To display the power of the projection method, sites are  randomly chosen in such a way that finite-ranged hopping amplitudes of the tight-binding Hamiltonian on the parent lattice do not couple them. Consequently, each site of the resulting amorphous lattice hosts localized atomic orbitals at the bare level, devoid of any topology. As we integrate or project out (from a path integral approach~\cite{Panigrahi2022}) the remaining sites of the parent crystal, the resulting renormalized Hamiltonian couples all the sites of the amorphous lattice through long-range hopping, featuring all the topological phases of the parent model. Constructed in this way, as the amorphous lattice emerges within a subsystem of the parent crystal, we call it {\it projected amorphous brane} (PAB).

\begin{figure}
    \centering
    \includegraphics[width=0.975\linewidth]{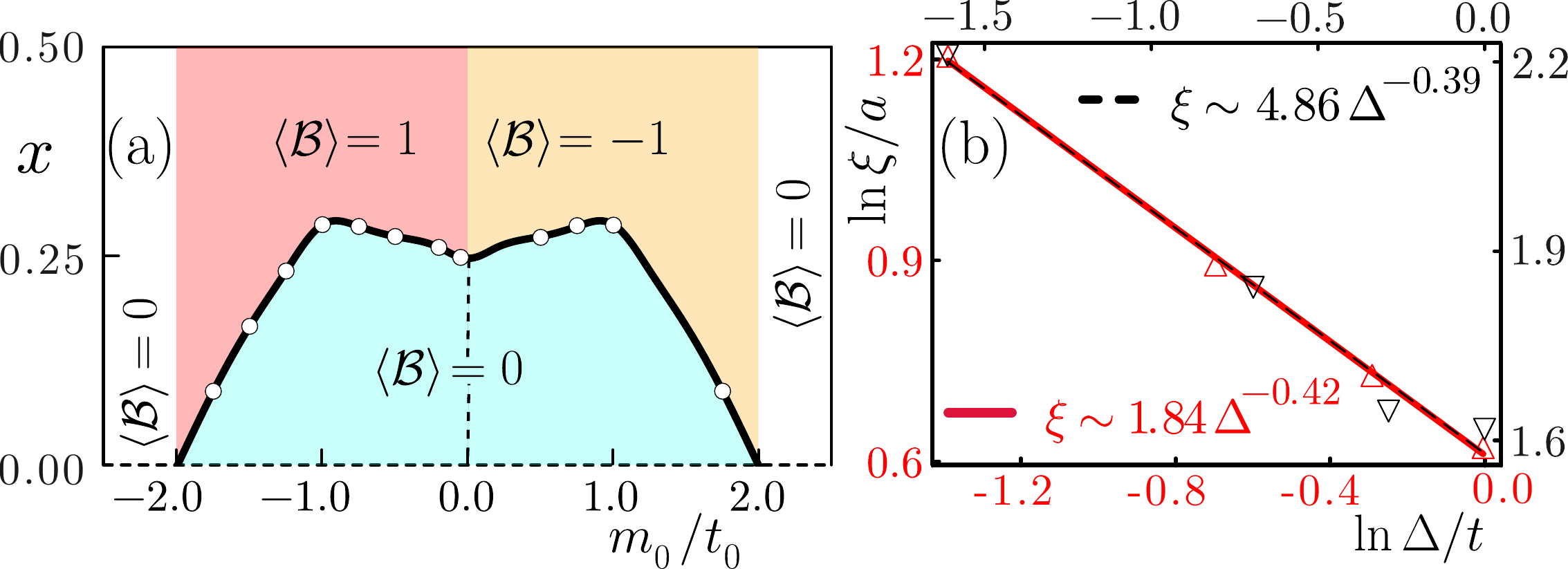}
    \caption{(a) Global phase diagram of Qi-Wu-Zhang model on PABs. Besides strong PATIs with the configuration-averaged BI $\langle {\mathcal B} \rangle=\pm 1$, the system fosters fragile PATIs within the topological regime ($|m_0/t_0|<2$) for smaller $x$ where $\langle {\mathcal B} \rangle=0$, but a small fraction of sites displays {\it almost} quantized LCM $ \approx {\mathcal B} $ (Fig.~\ref{fig:bulk-boundary}). White shaded regions are PANIs, where $ \langle {\mathcal B} \rangle$ and LCM vanish. Numerical calculations are performed down to $x=0.01$ and at circled points with the thick black curve extrapolating them. The horizontal dashed $x=0$ line is a singular limit, when the PAB becomes {\it empty}. (b) Dependence of average inter-site distance in the amorphous brane ($\xi/a$) on the half-spectral gap ($\Delta/t_0$) in a square lattice of lattice spacing $a$. Solid red (dashed black) line depicts $\xi = a/\sqrt{x}$ ($\xi = a/\sqrt{x-0.25}$) for which $\Delta = |2t_0+m_0|$ ($\Delta= | m_0|$) when $-2<m_0/t_0<-1$ ($-1<m_0/t_0<0$). We set $t=t_0=1$.
    }
    \label{fig:phase-scaling}
\end{figure}

\begin{figure*}
    \centering
    \includegraphics[width=0.9\linewidth]{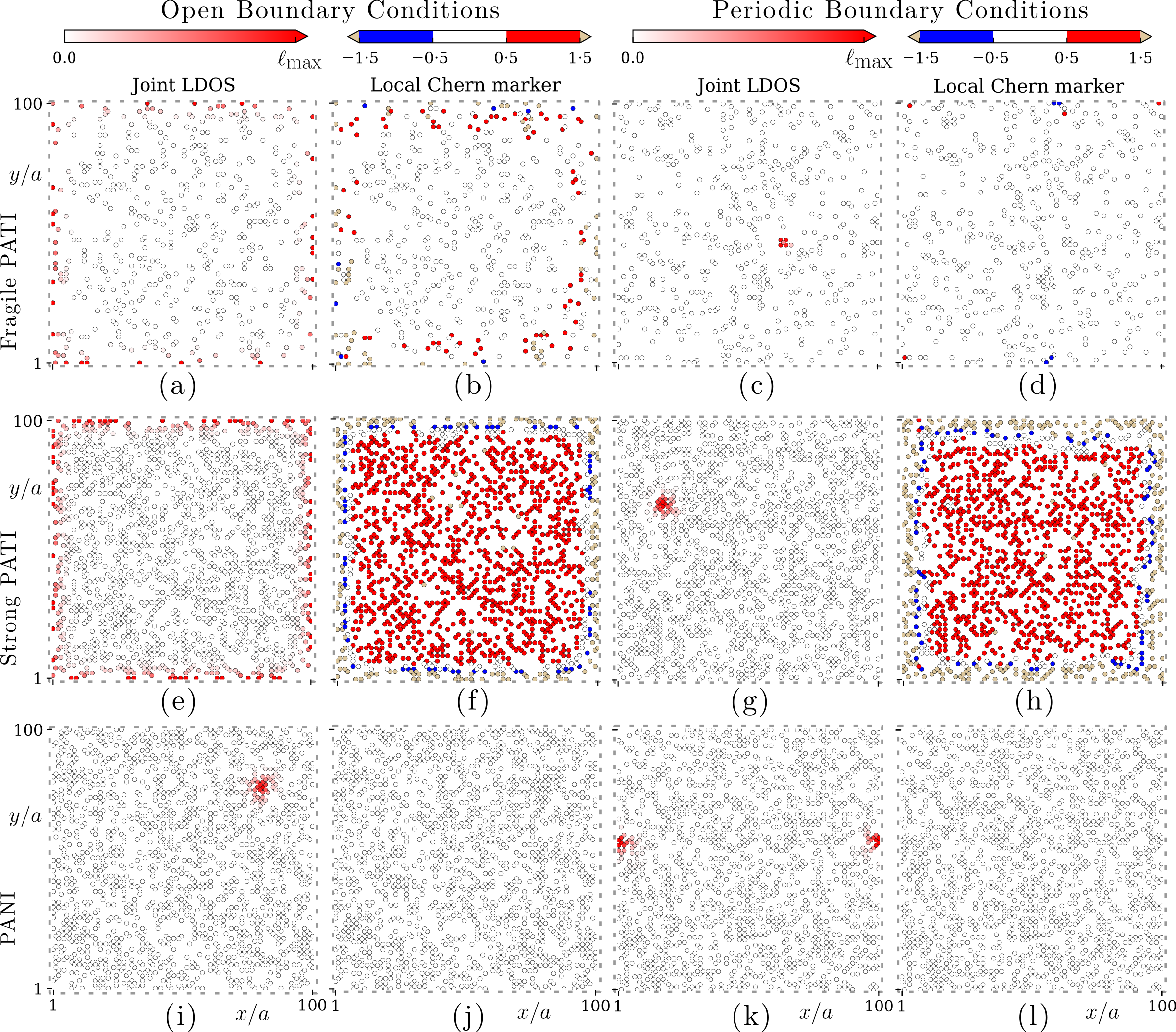}
    \caption{(a) Joint local density of states (LDOS) for closest to zero-energy modes with energies $\pm E_0$ for {\it fragile} PATIs with OBCs for $x=0.05$, $m_0=-1.75$, and $\ell_{\rm max} = 0.05$, where $E_0=0.168$. (b) LCM for the same configuration and parameters as in (a). Panels (c) [(d)] is analogous to (a) [(b)], but with PBCs. For (c) $E_0=1.957$ and $\ell_{\rm max} = 0.45$. Panels (e)-(h) are analogous to (a)-(d), respectively, but for {\it strong} PATIs and $x=0.15$. For (e) $\ell_{\rm max} = 0.015$ and $E_0= 0.055$, and for (g) $\ell_{\rm max} = 0.14$ and $E_0= 1.251$. Panels (i)-(l) are also analogous to (a)-(d), respectively, but for PANI with $m_0=-2.5$ and $x=0.15$. For (i) $\ell_{\rm max} = 0.15$ and $E_0= 1.426$, and for (k) $\ell_{\rm max} = 0.145$ and $E_0= 1.461$. We set $t=t_0=1$.   
    }
    \label{fig:bulk-boundary}
\end{figure*}

\begin{figure}[t!]
    \centering
    \includegraphics[width=1.00\linewidth]{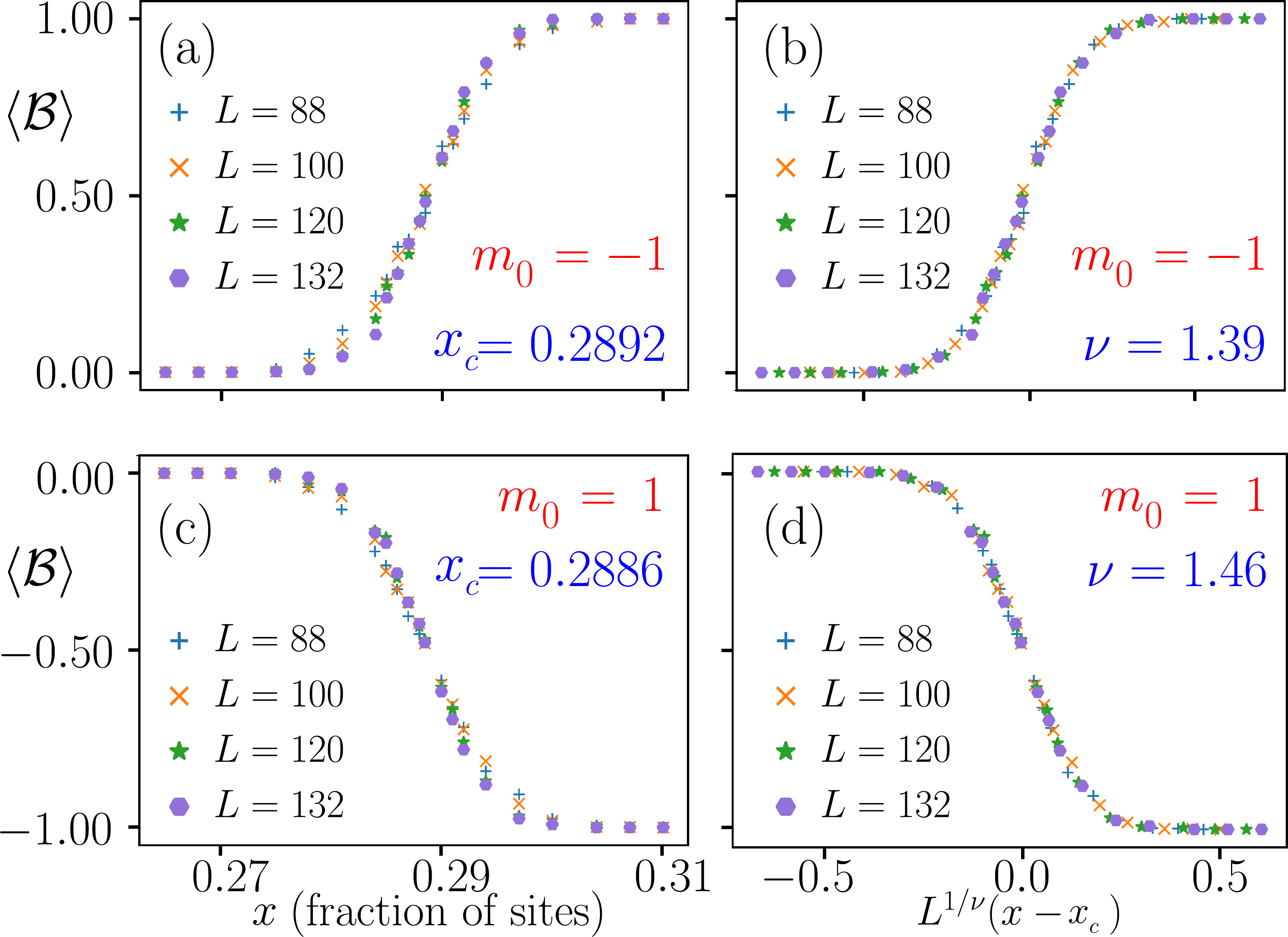}
    \caption{Configuration-averaged BI $\langle {\mathcal B} \rangle$ (from 800 independent realizations of the amorphous lattice) as a function of $x$ for (a) $m_0=-1$ and (c) $m_0=+1$ with $t=t_0=1$. Here, $x_c$ marks the critical concentration of sites, where $\langle {\mathcal B} \rangle \approx \pm 0.5$, for the QPT between strong and fragile PATIs, characterized by $\langle {\mathcal B} \rangle= \pm 1.0 $ and $\langle {\mathcal B} \rangle=0.0$, respectively. Data collapse, obtained by comparing $\langle {\mathcal B} \rangle$ with $L^{1/\nu} (x-x_c)$, yields the correlation length exponent (b) $\nu=1.39$ for $m_0=-1.0$ and (d) $\nu=1.46$ for $m_0=1.0$ for such transitions. See also Table~\ref{tab:values}. 
    }
    \label{fig:m=pm_1}
\end{figure}

Starting from a tight-binding Hamiltonian for time-reversal symmetry breaking insulators on a parent square lattice~\cite{Qi2006}, we showcase the following generic outcomes on its geometric descendant two-dimensional PABs. Within the entire topological parameter regime of the parent model [Fig.~\ref{fig:lattice}(b)], there always exists a critical concentration of sites ($x=x_c$) above which we find translationally active and inert {\it strong} PATIs. Such a phase is characterized by a quantized global topological invariant, namely the Bott index (BI) as well as a local topological marker, namely the local Chern marker (LCM), on a large fraction of sites therein. Within the same parameter regime, however, for $x<x_c$, we identify a peculiar insulating phase on PABs for which the global topological invariant vanishes, but a small fraction of sites continues to manifest almost quantized local topological marker. We name this phase a {\it fragile} PATI. See Fig.~\ref{fig:phase-scaling}(a). At $x=x_c$, the average inter-site distance ($\xi$) displays a {\it universal} scaling with the band gap of parent model ($\Delta$) [Fig.~\ref{fig:phase-scaling}(b)]. Both strong and fragile PATIs display boundary-localized in-gap modes in systems with open boundary conditions (OBCs). We also find projected amorphous normal insulators (PANIs) in the entire trivial parameter regime of the parent model for any $x$. Both global and local topological invariants vanish for PANIs, devoid of any boundary-localized in-gap mode. See Fig.~\ref{fig:bulk-boundary}. Across the strong-to-fragile PATI quantum phase transition (QPT) at $x=x_c$, the configuration-averaged global topological invariant displays a single parameter scaling. The {\it mean} value of the associated correlation length exponent $\nu \in (1.00,1.46)$, which is possibly dimension-dependent, but not on its symmetry class, as the QPT is tuned structurally by $x$. See Fig.~\ref{fig:m=pm_1} and Table~\ref{tab:values}.

{\it Projected brane}.~As the method of projection lies at the heart of this work, we discuss it in detail~\cite{Panigrahi2022}. We denote the segment of the parent tight-binding Hamiltonian (including hopping and on-site terms) operative solely among the sites constituting the amorphous lattice [red colored sites in Fig.~\ref{fig:lattice}(a)] as $H_{11}$ and that among the sites falling outside such a brane [black colored sites in Fig.~\ref{fig:lattice}(a)] as $H_{22}$. Elements of the hopping Hamiltonian between these two sets of sites are denoted by $H_{12}$ and $H_{21}$ ($ \equiv H^\dagger_{12}$). To integrate out the sites falling outside the amorphous brane, we write the imaginary-time ($\tau$) partition function (${\mathcal Z}$) in terms of the Grassmann variables $\Psi_1$ and $\bar{\Psi}_1$ (operative on the sites of the amorphous lattice), and $\Psi_2$ and $\bar{\Psi}_2$ (acting on the remaining sites of the parent crystal) as
\begin{eqnarray}
    {\mathcal Z} = \int \mathcal{D} \Psi_1 \mathcal{D} \Psi_2 \mathcal{D} \bar{\Psi}_1 \mathcal{D} \bar{\Psi}_2 \: \exp{-S[\Psi_1,\Psi_2, \bar{\Psi}_1, \bar{\Psi}_2]}
\end{eqnarray}
with the action $S[\Psi_1,\Psi_2, \bar{\Psi}_1, \bar{\Psi}_2] \equiv S$ given by
\begin{eqnarray}
    S = \int_0^\beta d\tau 
    \begin{pmatrix}
    \bar\Psi^{\tau}_1 & \bar\Psi^{\tau}_2
    \end{pmatrix}
    \begin{pmatrix}
        \partial_\tau + H_{11} & H_{12}\\
        H_{21} & \partial_\tau + H_{22}
    \end{pmatrix} \begin{pmatrix}
    \Psi^{\tau}_1 \\ \Psi^{\tau}_2
    \end{pmatrix},
\end{eqnarray}
where $\beta$ is the inverse temperature. In terms of the Grassmann variables in the frequency space $\Psi^\tau_{1,2} = \sum_{\omega_n} c^{(n)}_{1,2} e^{-i\omega_n \tau}/\sqrt{\beta}$, where $\omega_n = (2n + 1) \pi /\beta$ with $n \in {\mathbb Z}$ (set of integers) are the fermionic Matsubara frequencies 
\begin{widetext}
\allowdisplaybreaks[4]
\begin{equation}
\begin{aligned}    
    S &= \sum_{\omega_n} \bar c^{(n)}_{1} (-i\omega_n + H_{11} ) c^{(n)}_{1} + \bar c^{(n)}_{1} H_{12}  c^{(n)}_{2} + \bar c^{(n)}_{2} H_{21} c^{(n)}_{1} + \bar c^{(n)}_{2} (-i\omega_n + H_{22} ) c^{(n)}_{2} \\
    & = \sum_{\omega_n} \bar c^{(n)}_{1} (-i\omega_n + H_{11} ) c^{(n)}_{1} - \bar c^{(n)}_{1} (H_{12} (-i\omega_n + H_{22})^{-1} H_{21})  c^{(n)}_{1} + \bar d^{(n)}_{2} (-i\omega_n + H_{22}) d^{(n)}_{2}.
\end{aligned}
\end{equation}
After integrating out the shifted variables $d^{(n)}_{2} = c^{(n)}_{2} + (-i\omega_n + H_{22})^{-1} H_{21} c^{(n)}_{1}$, we arrive at the effective action
\begin{equation}
\begin{aligned}
    S_{\rm eff}[\Psi_1, \bar \Psi_1] = \sum_{\omega_n} \bar c^{(n)}_{1} [-i\omega_n + H_{11}
     - H_{12} (-i\omega_n + H_{22})^{-1} H_{21}]  c^{(n)}_{1}
\end{aligned}
\end{equation}
\end{widetext}
for the sites on the amorphous brane. From $S_{\rm eff}[\Psi_1, \bar \Psi_1] $ we obtain an effective Hamiltonian ($H_{\rm eff}$) that describes renormalized hopping elements therein after taking the limit $\omega_n \to 0$, in terms of sharp quasiparticle excitations
\allowdisplaybreaks[4]
\begin{equation}~\label{eq:Heff}
    H_{\rm eff} = H_{11} - H_{12} H_{22}^{-1} H_{21}.
\end{equation}
We arrive at all the conclusions by diagonalizing $H_{\rm eff}$.

{\it Microscopic model}.~Although the derivation of $H_{\rm eff}$ is insensitive to the dimensionality, symmetry class, and underlying parent lattice structure, for concrete demonstrations of translationally active and inert, strong and fragile PATIs, and PANIs we consider the Qi-Wu-Zhang model for time-reversal symmetry breaking insulators on a square lattice of lattice spacing $a$~\cite{Qi2006}. The corresponding Bloch Hamiltonian $H(\vec{k})={\boldsymbol \tau} \cdot \vec{d}(\vec{k})$, where $\vec{k}$ is the lattice momentum, ${\boldsymbol \tau}$ is the vector Pauli matrix, and $d_1(\vec{k})=t \sin(k_x a)$, $d_2(\vec{k})=t \sin(k_y a)$, and $d_3(\vec{k})=m_0-t_0 [\cos(k_x a)+\cos(k_y a)]$. This model supports topological (normal) insulators for $|m_0/t_0|<2$ ($|m_0/t_0|>2$). The topological regime fragments into two sectors. The translationally inert (active) $\Gamma$ (${\rm M}$) phase is realized for $0<m_0/t_0<2$ ($-2<m_0/t_0<0$). The topological invariant first Chern number~\cite{Thouless1982}, ${\mathcal C}=-1,1$, and $0$ for the $\Gamma$ phase, ${\rm M}$ phase, and normal insulators, respectively, with ${\mathcal C} \equiv {\mathcal B}$. We set the hopping amplitudes $t=t_0=1$. The QPT between topological and normal insulators through band gap closing at the ${\rm M}$ ($\Gamma$) point occurs when $m_0=-2$ ($m_0=2$), and that between the translationally active and inert phases occurs via a band gap closing at the ${\rm X}$ and ${\rm Y}$ points when $m_0=0$. Thus, the system falls in the basin of attraction of the ${\rm M}$ ($\Gamma$) point for $-2 \leq m_0 \leq -1$ ($1 \leq m_0 \leq 2$) , and the ${\rm X}$ and ${\rm Y}$ points for $-1 \leq m_0 \leq 1$.

The Bloch Hamiltonian can readily be implemented on a square lattice using Fourier transformations, from which we obtain $H_{\rm eff}$ for PABs [Eq.~\eqref{eq:Heff}]. The sites constituting the amorphous lattice are randomly chosen from the square lattice in such a way that they are not connected by the hopping terms proportional to $t$ and $t_0$. Thus the bare Hamiltonian for an amorphous brane $H_{11}$ describes a {\it trivial} insulator composed of site-localized {\it atomic} orbitals at energies $\pm m_0$. By contrast, the second term in $H_{\rm eff}$, obtained via projection, introduces longer range hopping which in turn endows PABs with topology. In this context, our findings are the following.

{\it Results}.~To identify topologically distinct phases and QPTs among them, we primarily compute the BI (${\mathcal B}$), ideally suited for systems without any crystalline order. The projector to the occupied single-particle states is
\begin{equation}
    P = \sum_{E_n < 0} \ket{n}\bra{n},
\end{equation}
where $\ket{n}$ satisfies $H_{\rm eff} \ket{n}=E_n \ket{n}$ and its complement to the unoccupied states is $Q = \mathbf{I} - P$ with $\mathbf{I}$ as the identity matrix. The BI is defined as~\cite{Loring2010} 
\begin{equation}
    \mathcal{B} =\frac{1}{2\pi}\Im \left[\tr \left[\log[V_y U_x V_y^\dagger U_x^\dagger]\right]\right],
\end{equation}
where $U_x = P U_0 P + Q$ with $U_0 = \exp[i 2 \pi X/L]$, $V_y = P V_0 P + Q$ with $V_0 = \exp[i 2\pi Y/L]$, and $X$ ($Y$) is the position operator, containing $x$ ($y$) coordinates of each site of the amorphous brane in the frame of the square lattice with periodic boundary conditions (PBCs) and linear dimension of $L$ in each direction.

In terms of the configuration-averaged BI $\langle {\mathcal B} \rangle$, we identify translationally active (inert) strong PATIs with $\langle {\mathcal B} \rangle=1$ ($\langle {\mathcal B} \rangle=-1$) for $-2<m_0 <0$  ($0<m_0 <2$) above a critical concentration of lattice sites on PABs $x_c$, which depends on $m_0$. For $|m_0|>2$, we find PANIs with ${\mathcal B} =0$ for any $x$. On the other hand, when $|m_0|<2$ (topological regime) but $x<x_c$, there exists an insulating phase where $\langle {\mathcal B} \rangle=0$. See Fig.~\ref{fig:phase-scaling}(a). To shed light on such an insulating phase, we compute the LCM on each site at $\vec{r}$ of the PAB, defined as~\cite{Bianco2011, Manna2024}
\begin{equation}
    \mathcal C(\vec r) = \frac{4\pi}{A} \Im \big( \bra{\vec r}  \tr \left[P X Q Y P\right]\ket{\vec r} \big),
\end{equation}
where `tr' implies sum over on-site orbitals, and $A = a^2/x$ is the average area per site therein. Remarkably, despite $\langle {\mathcal B} \rangle=0$, a finite (although small) fraction of sites of the PAB features almost quantized LCM  therein, peaked around ${\mathcal B}$ found in the parent square lattice, especially when we impose OBCs. The number of such topological sites decreases drastically with PBCs. But, it should be noted that local topological markers are strictly defined with OBCs. We name this phase fragile PATI. Below $x_c$, as $x$ increases the number of sites showing quantized LCM also increases~\cite{SM_PABs}. The LCM is quantized to $\langle {\mathcal B} \rangle$ (approximately) on a majority of sites in strong PATIs, whereas it vanishes everywhere in the system in PANIs irrespective of the boundary conditions. See Fig.~\ref{fig:bulk-boundary}.

\begin{table}[t]
    \centering
    \begin{tabular}{|c|c|c|}
        \hline
        $m_0/t_0$ & $x_c$ & $\nu$ \\
        \hline
        $- \; 1.75$ & $0.0898$ & $1.27 \pm 0.26$ \\
        $- \; 1.50$ & $0.1675$ & $1.00 \pm 0.13$ \\
        $- \; 1.25$ & $0.2337$ & $1.15 \pm 0.16$ \\
        $- \; 1.00$ ($+ \; 1.00$) & $0.2892$ ($0.2886$) & $1.39 \pm 0.25$ ($1.46 \pm 0.26$) \\
        \hline
    \end{tabular}
    \caption{Critical site concentration on PABs ($x_c$) for the strong-to-fragile PATI QPT and the corresponding correlation length exponent ($\nu$) for various $m_0$ when $t=t_0=1$~\cite{SM_PABs}.}
    \label{tab:values}
\end{table}

Quantized BI or LCM corroborates the bulk-boundary correspondence through edge-localized in-gap modes visible on PABs with OBCs, whose energies are at least an order of magnitude smaller than the bulk gap in systems with PBCs. Such modes are only found in the strong and fragile PATIs; not in PANIs. See Fig.~\ref{fig:bulk-boundary}. We also confirm that the edge modes of strong and fragile PATIs are robust against sufficiently weak random pointlike charge impurities up to a strength of 15\% of the bandwidth, computed on the parent square lattice~\cite{SM_PABs}.

At $x=x_c$, where the strong-to-fragile PATI QPT takes place, the average distance between the sites ($\xi/a$) follows a universal scaling with $\Delta/t_0$. Specifically, when the system falls in the basin of attraction of the ${\rm M}$ point ($-2 \leq m_0 \leq -1$), $\xi \sim \Delta^{-0.42}$ with $\xi \sim 1/\sqrt{x}$ and $\Delta=|2t_0+m_0|$. By contrast, when the system falls in the basin of attraction of the ${\rm X}$ and ${\rm Y}$ points ($-1<m_0/t_0<0$), $\xi \sim \Delta^{-0.39}$ with $\xi \sim 1/\sqrt{x_\star-x}$, $\Delta=|m_0|$, and $x_\star = 0.25$. See Fig.~\ref{fig:phase-scaling}(b). A finite value of $x_\star$ suggests that despite being randomly distributed, the collection of such lattice sites on PABs develops randomness over an effective uniform background of lattice sites with density $a/\sqrt{x_\star}$. Notice that in both cases the power-law dependence of $\xi$ on $\Delta$ is almost {\it equal}, suggesting a {\it universal} scaling between them.

Finally, we note that across the strong-to-fragile PATI QPT, $\langle {\mathcal B} \rangle$ displays a single parameter scaling. The critical site concentration ($x_c$) for such a QPT is pinned at an $x$, where $\langle {\mathcal B} \rangle$ on PABs for different parent square lattices with varying $L$ roughly cross each other and $\langle {\mathcal B} \rangle \approx \pm 0.5$. With the knowledge of $x_c$, when $\langle {\mathcal B} \rangle$ is compared with $L^{1/\nu} (x-x_c)$ we observe excellent data collapses for a correlation length exponent $\nu$, as shown in Fig.~\ref{fig:m=pm_1} for $m_0= \pm 1$. The values of $\nu$ for various choices of $m_0$ are reported in Table~\ref{tab:values}. Numerically obtained {\it mean} values of $\nu$ for structural quantum Hall plateau transition range between $1.00$ and $1.46$~\cite{SM_PABs}, which is distinct from the one when such a QPT is triggered by on-site disorder for which $\nu \in (2.3,2.6)$~\cite{Chalker1988, Slevin2009, Sbierski2021, Salib2025}. The errorbar of $\nu$ is determined using the {\it bootstrap method}, yielding the standard error. Subsequently, the $95\%$ confidence interval is estimated to be $3.92$ times the standard error, assuming the sampling distribution is approximately normal~\cite{SM_PABs}.

{\it Summary and discussions}.~To summarize, here we introduce two-dimensional PABs, constituted by a fraction of sites ($x$), selected randomly from the parent square lattice, such that they are disconnected by any hopping amplitudes at the bare level. The effective Hamiltonian for such PABs, constructed by integrating out sites of the parent crystal falling outside it, captures the footprints of both translationally active and inert topological phases with quantized global and local topological invariants, a phase of matter named strong PATI, besides the PANI for which both types of topological invariant vanish. More intriguingly, at lower concentrations of sites ($x<x_c$) we unearth a peculiar topological phase, for which the global topological invariant vanishes, but the local one displays almost quantized values; a phase of matter we coined fragile PATI. Both strong and fragile PATIs accommodate in-gap edge-localized modes, corroborating the bulk-boundary correspondence. Together, the present results illustrate a universal aspect of {\it projected topological branes} through which one captures a wide variety of lattice-based topological phases on their geometric descendant lower-dimensional quasicrystals and emergent crystals~\cite{Panigrahi2022, Tyner2024, Panigrahi2026}, and fractal~\cite{Salib2024} and amorphous lattices, devoid of some or all the crystalline symmetries as well as emergent non-Hermitian topology~\cite{Dordevic2026} and disorder-driven QPTs~\cite{Tyner2025} therein. It will be worthwhile extending the construction of PABs to harness crystalline, weak, and higher-order topological insulators and superconductors in various dimensions, belonging the different symmetry classes, which we leave for forthcoming investigations. Given that local topological markers can be defined for all Altland-Zirnbauer symmetry classes~\cite{Chen2023}, we believe that fragile amorphous topological phases can also be recognized there.

The requisite long-range hopping, generated during the projection, through which otherwise randomly selected disconnected sites constituting the PAB get connected and in turn foster all the parent crystal-based topological phases, however, makes it challenging to realize strong and fragile PATIs in quantum materials immediately. Still, recent progress with designer quantum materials makes us optimistic regarding their experimental realizations in laboratory~\cite{Gomes2012, Collins2017, Kempkes2019, Ma2025, Yang2025}. Nonetheless, various highly tunable classical metamaterials, such as topolectric circuits~\cite{Imhof2018, Dong2021, Olekhno2022}, and photonic~\cite{Ozawa2019} and phononic~\cite{Susstrunk2015, Yang2015} lattices are fascinatingly promising to realize all the theoretically predicted PATIs in the near future.

{\it Data availability}.~Numerical code and data generated during this work are available in Ref.~\cite{Panigrahi_PAB_codes}.

{\it Acknowledgments}.~Majority of the computation was carried out with resources provided by subMIT at MIT Physics~\cite{MITSubmit}. A.P.\ thanks Minho Luke Kim and Vladislav Poliakov for technical support. B.R.\ was supported by NSF CAREER Grant No.\ DMR-2238679. We thank Vladimir Juri\v ci\' c for comments on the manuscript.

\end{document}